\documentclass{article}
\usepackage{amsmath}
\usepackage{amssymb}
\usepackage{graphicx}
\usepackage{float}
\usepackage [left=1in,right=1in,top=1in,bottom=1in] {geometry}

\begin{document}
\everymath{\displaystyle}
\title{Gamma scattering scanning of concrete block for detection of voids.}
\author{Shivaramu$^{1}$, Arijit Bose$^{2}$ and M. Margret$^{1}$ \\
$^{1}$Radiological Safety Division, Safety Group, IGCAR, Kalpakaam - 603 102 (India) \\
$^{2}$ Chennai Mathematical Institute, Sipcot I.T. Park, Chennai - 603 103 (India)}
\date{E-mail: shiv@igcar.gov.in, arijitbose@cmi.ac.in}
\maketitle

\begin{center}
\title{\textbf{Abstract}}
\end{center}
The present paper discusses a Non Destructive Evaluation (NDE) technique involving Compton back-scattering. Two 15 cm x 15 cm x 15 cm cubical concrete blocks were scanned for detection of voids. The setup used a PC controlled gamma scattering scanning system. A $^{137}$Cs radioactive source of strength 153.92 GBq with lead shielding and a collimated and shielded 50\% efficiency coaxial HPGe detector providing high resolution energy dispersive analysis of the scattered spectrum, were mounted on source and detector sub-assemblies respectively. In one of the concrete blocks air cavities were created by insertion of two hollow cylindrical plastic voids each of volume 71.6 cm$^3$. Both the concrete blocks, one normal and another with air cavities were scanned by lateral and depth-wise motion in steps of 2.5 cm. The results show that the scattering method is highly sensitive to changes in electronic and physical densities of the volume element (VOXEL) under study. The voids have been detected with a statistical accuracy of better than 0.1\% and their positions have been determined with good spatial resolution. A reconstruction algorithm is developed for characterization of the block with voids. This algorithm can be generally applied to such back-scattering experiments in order to determine anomalies in materials.\\  

\section{\textbf{Introduction}}
The need for advanced techniques for detection and evaluation of a class of sub-surface defects that require access only to the one side of any material or structure relatively thick to be inspected has drawn attention to X-ray or gamma backscatter as a desirable choice [1-3].There is a great demand for non-destructive testing and evaluation (NDT, NDE) techniques in many diverse fields of activity. A particular area of interest is encountered in detection of defects in concrete structures. Ultrasonic, Radiography and Compton Back Scattering techniques are currently employed to test the components of structures. Transmission provides line-integrated information along the path of radiation from the source to the detector, which masks the position of an anomaly present along the transmission line. Therefore, it is difficult to determine the position of an anomaly directly from transmission measurements. The backscatter technique can image a volume rather than a plane. Point-wise information can be obtained by focusing the field of view of the source and detector so that they intersect around a point. Since both the source and detector are located on the same side of the object, examination of massive or extended structures become possible. This NDE technique should be developed as it can be applied in:
\begin{itemize}
	\item{Detection of corrosion in steel liners that are buried (covered by concrete), detection of voids $>$ 20 mm diameter in concrete. Detection of flaws before they propagate to the point of causing failure is essential.}
  \item{Locating flaws in nuclear reactor walls or dam walls.} 
	\item{Detection of landmines.}
\end{itemize}
  Therefore an urgent need exists to develop diversified and effective NDE technologies to detect flaws in structures. The gamma scattering method is a viable tool for inspecting materials since it is strongly dependent on the electron density of the scattering medium, and in turn, its mass density. The concept is based on the Compton interaction between the incident photons and the electrons of matter. Gamma or X-rays scattered from a VOXEL are detected by a well-collimated detector placed at an angle which could vary from forward scattering angles to the back-scattering configuration. The scattered signal, therefore, provides an indication of the electron density of the material comprising the inspected volume [4]. By scanning a plane of interest of the object, it is possible to obtain the density distribution in this plane. In this process scattered signals are recorded from different depths of the material. Hence, gamma scattering enables the detection of local defects and the discrimination between materials of different density and composition, such as concrete, void and steel. Moreover, in NDT, the nature of the inspected object is usually known and the purpose is to determine any disturbance in the measured signal that can indicate the presence of an anomaly.\\ 

Two concrete blocks one normal and another with air cavities were scanned completely by voxel method. The inspection volume (voxel) was formed by the intersection of the collimated incident beam cone and the collimated field of view of the HPGe detector. The corresponding Compton scattered spectrums of these two blocks were compared and analyzed. As the void intersected the sensitive volume, there was a decrease in the total electron density of the material comprising the voxel, hence it showed a decrease in detector response. The presence of void could be clearly distinguished and located. 

\section{\textbf{Experimental Procedure}}
The schematic of the experimental set up is shown in Fig.1. The scanning system consists of source and detector unit and a four axis job positioning system for moving the block.\\
 
\begin{figure}[H]
\begin{center}
\includegraphics[scale=0.7]{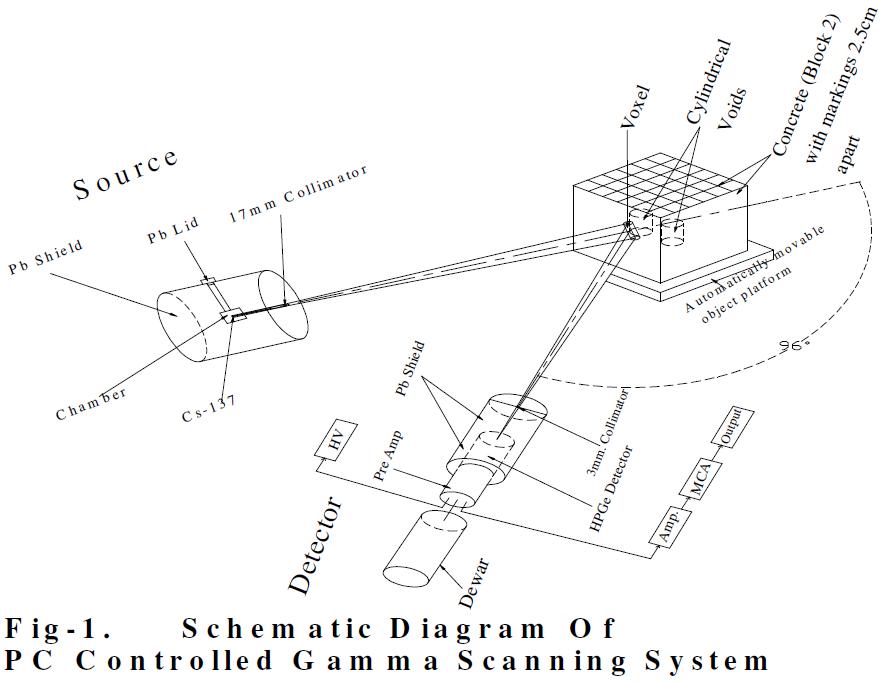}
\end{center}
\end{figure}

The source and detector units are composed of a positioning stage, a fabricated base and common to both unit is a control panel and a Laptop. The positioning stages of each unit consist of X,Y-axis travel stage, Z- axis vertical travel stage and a rotary stage. The $^{137}$Cs radioactive source of strength 153.92 GBq with a lead shielding and the collimated and shielded 50\% efficiency coaxial HPGe detector, providing high resolution energy dispersive analysis of the scattered spectrum, are mounted separately on the source and detector sub-assemblies of 6-axis system respectively. The concrete blocks are mounted on the 4-axis job positioning system. The voxel to be analyzed is geometrically established by the intersection of the incident and scattered beams. The size of the voxel is defined by the diameter and length of the collimators employed and on source to object, detector to object distances and can be easily chosen by proper adjustment of X, Y, Z and $\theta$ positions of 6-axis and 4-axis job positioning systems. The source and detector is collimated with solid lead cylindrical collimators of diameter 17 and 3 mm respectively, and the size of the resulting voxel is 19 cm$^3$. The scattered intensity from a voxel in the concrete block is detected and the pulse height spectrum (PHS) is accumulated and displayed using an 8K-channel analyzer which is interfaced with a PC for data storage and analysis. The scattering angle is $96^\circ$ and the incident photon energy emitted by $^{137}$Cs is 661.6 keV and the energy of the scattered photon is 273.6 keV (Compton scattering). Two concrete blocks one normal and another with air cavities, created by inserting two hollow cylindrical plastic voids of diameter 39 mm and height 60 mm, are chosen for void detection and quantification. These voids were placed at equal distances from the surface and from the centre of concrete block symmetrically as shown in Fig.1. The horizontal and depth-wise scanning of a plane in the concrete cube was performed by moving the blocks across the source and detector collimators in steps of 2.5 cm. The photo peak counts of the scattered spectrum for corresponding positions of both the specimens of concrete blocks one normal and another with air cavities were compared and analyzed.

\section{\textbf{Reconstruction Algorithm and Attenuation Corrections}}
The reconstruction algorithm developed to correct for attenuation of the incident and scattered photons by the material surrounding each scatter site is described. It provides accurate reconstruction of the Compton scattering data through an adequate correction of the absorption phenomena. The path from the source to detector can be broken into three stages. The first stage is the photon's travel from the source to the scattering point P along the path $\alpha$. Neglecting attenuation due to air, from Beer-Bougher law:
\begin{equation}I_1=I_0 \text{exp}\left[-\int_{0}^{x}\left\{\frac{\mu(E)}{\rho}\right\}\rho dx'\right]\end{equation}
Where $I_1$ and $I_0$ are the transmitted and incident flux, respectively, $\frac{\mu(E)}{\rho}$is the mass attenuation coefficient of the material for photons of energy $E$, $\rho$ is the density of the material and x is the length of path $\alpha$ in block [5].\\
The second stage is scattering towards detector at point P. The scattered flux $I_2$ is determined by;
\begin{equation}I_2=I_1\frac{d\sigma}{d\Omega}(E,\theta)S(E,\theta,Z)d\Omega\rho_e(P)\Delta L \end{equation}
\begin{figure}[H]
\begin{center}
\includegraphics[scale=0.7]{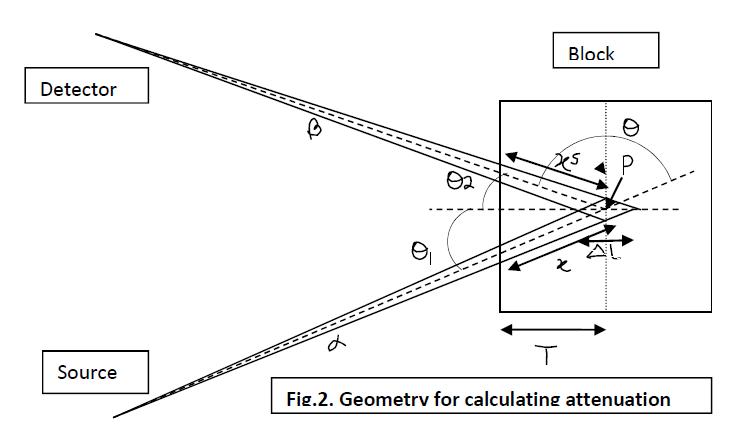}
\end{center}
\end{figure}
Where $\frac{d\sigma}{d\Omega}(E,\theta)$ is the differential scatter cross section as governed by the Klein-Nishina formula (a function of the incident gamma energy $E$ and scatter angle $\theta$ ), $S$ is the incoherent scattering function (a function of $E, \theta$ and atomic number of the element $Z$), $d\Omega$ is the solid angle subtended by the detector and its collimator, $\rho_e(P)$ is the electron density at point $P$ and $\Delta L$ is an element of path $\alpha$ in the vicinity of point $P$ (the voxel thickness). The electron density at $P$ is the material property we are attempting to measure. It is proportional to the physical density $\rho$ according to the formula: $\rho_e =\frac{\rho NZ}{A}$ [6], where $N$ is Avogadro's number, $Z$ is atomic number and $A$ is atomic weight. The third stage is the transport of scattered photons back through material to detector. The signal is further attenuated, so that:
\begin{equation}I_3=I_2 \text{exp}\left[-\int_{0}^{x^s}\left\{\frac{\mu ^s(E^s,\theta)}{\rho}\right\} \rho dx'\right] \end{equation}
here $I_3$ is the flux intensity reaching the detector, $\frac{\mu^s(E^s,\theta)}{\rho}$ is the mass attenuation coefficient for scattered photons of energy $E^s$, now a function of $\theta$ by the virtue of the Compton energy shift at $P$, and $x^s$ is the length of the path $\beta$ in the block. Combining the expressions for the three stages, the signal intensity corresponding to point $P$ can be written as, 
\begin{equation}I_3=I_0 \text{exp}\left[-\int_{0}^{x}\left\{\frac{\mu(E)}{\rho}\right\}\rho dx' + \int_{0}^{x^s}\left\{\frac{\mu ^s(E^s,\theta)}{\rho}\right\} \rho dx'\right]\frac{d\sigma}{d\Omega}(E,\theta)S(E,\theta,Z)d\Omega\rho_e(P)\Delta L \end{equation}
Let \begin{equation} k=\frac{d\sigma}{d\Omega}(E,\theta)S(E,\theta,Z)d\Omega \frac{NZ}{A}\Delta L \end{equation}
The attenuation factor is,
\begin{equation} AF=exp-[\int_{0}^{x}\{\frac{\mu(E)}{\rho}\}\rho dx' + \int_{0}^{x^s}\{\frac{\mu ^s(E^s,\theta)}{\rho}\} \rho dx'] \end{equation}
In the present case as per Fig.2;
\begin{equation} AF=exp-[\int_0^T\{\frac{\mu(E)}{\rho}\}\rho\frac{dt}{cos\theta_1}+\int_0^T\{\frac{\mu^s(E^s)}{\rho}\}\rho\frac{dt^s}{cos\theta_2} \end{equation}
combining (4), (5) and (6) we get,
\begin{equation} I_3=I_0k\rho(P) AF \end{equation}
Taking the ratio of $I_3$ which is experimentally the counts under the photo peak of Compton scattered spectra for normal concrete block (Block 1) and another with air cavities (Block 2) we get,
\begin{equation} \frac{I_3(Block2)}{I_3(Block1)}=\frac{\rho(P, Block2) AF(P, Block2)}{\rho(P, Block1) AF(P, Block1)} \end{equation}
The attenuation factor for normal concrete block can be calculated just by substituting the mass attenuation factor and density of concrete in (7). For calculating the attenuation factor for different VOXEL positions of concrete block with air cavities one requires the geometry and hence the path length traveled by incident and scattered rays in void and concrete. This information is obtained using software graphics. Here the knowledge of the precise location and size of the voids are used.
   
We see that equation (9) is iteration in density. We can either take up iteration approximations to obtain the void locations in case of voids of unknown location and sizes. Other alternatives are:
\begin{itemize}
	\item {reduction of VOXEL size which will give a better spatial resolution and contour of the voids}
	\item {or changing the scattering angle and generating more density distribution pictures of the concrete block under investigation and finally superimposing the results.}
\end{itemize}
\section{\textbf{Results and Discussion}}
The lateral variation of the difference in Compton photo peak counts of the scattered spectrum between normal and void incorporated concrete block, measured at 7.5 cm from the bottom, for various depths (D0:-front face to D3:-center each 2.5 cm apart) are shown in Fig.3 and 4. The density ratio of the two concrete blocks derived from the equation (9) plotted as a function of lateral distance is shown in Fig.5. As the sensitive volume intersects the void cavity, there is a reduction in the total electron density and hence increase in the difference scattered intensity (Figs. 3 \& 4) and decrease in density ratio (Fig.5). At the front face of the blocks (D0) it is known that there is no void coming into the investigation voxel and the same can be inferred from the results shown in Figs.3, 4 \& 5 which show same scattered intensity and hence density. On the other hand at 5 cm depth from the front face (D2) the effect of the voids can be seen (Figs.3 \& 4) as a U shaped scattered intensity curve and an inverted U shaped density ratio (Fig.5). The magnitude of the difference in scattered intensity (Figs.3 \& 4) and decrease in density (Fig.5) is proportional to the size of the void within the voxel. This gives an idea of the size and location of the void as the voxel size and its orientation in concrete is known for all positions. The voxel size was estimated as 85 cm$^{3}$ in the present case. The cylindrical voids have been detected in the present study with a statistical accuracy of better than 0.1\%. It is also possible to locate the position of the voids successfully even without considering the contribution due to the attenuation factor. The effectiveness of this inspection technique can be defined by the spatial resolution and the density contrast achievable. Good resolution requires a small sensitive volume, while high contrast demands large sensitivity to changes in composition. The size of the inspection volume defines the spatial resolution. Reducing the collimator aperture to improve the spatial resolution leads, however, to a decrease in the count rate. This can be compensated for by increasing the source strength and the counting period. A practical compromise is therefore necessary to achieve a reasonable resolution within an appropriate counting period and without exposure to a high dose of radiation. In order to increase the contrast, the contribution to the detector from the material contained within the sensitive volume should be enhanced while that of the surrounding media should be reduced. This can be achieved by reducing the attenuation of the radiation as it travels to and from the sensitive volume, and/or by increasing the probability of scattering within the volume. The attenuation and scattering probability, however, depend on the radiation energy and the angle of scattering. The incident angle, defined in Fig. 2. determines the photon path length and in turn affects the attenuation probability. The source energy and the scattering angle are, however, the two most important design parameters as they directly affect the detector response. 
\begin{figure}[H]
\begin{center}
\includegraphics[scale=0.7]{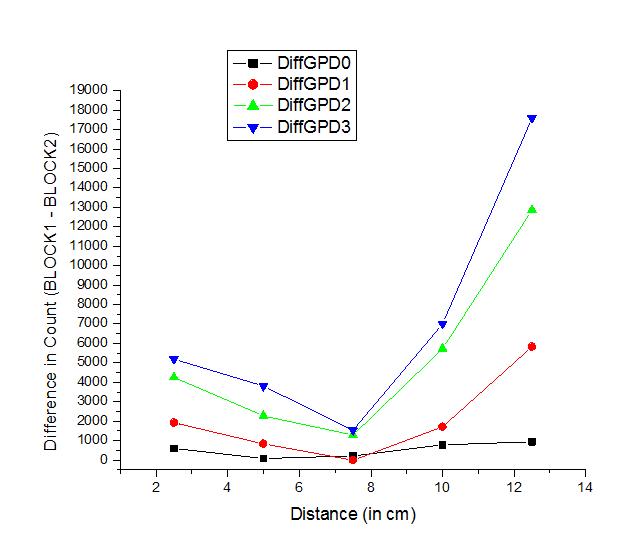}
\mbox{\textbf{Fig.3.} The difference Compton photo peak counts}
\end{center}
\end{figure}

\begin{figure}[H]
\begin{center}
\includegraphics[scale=0.7]{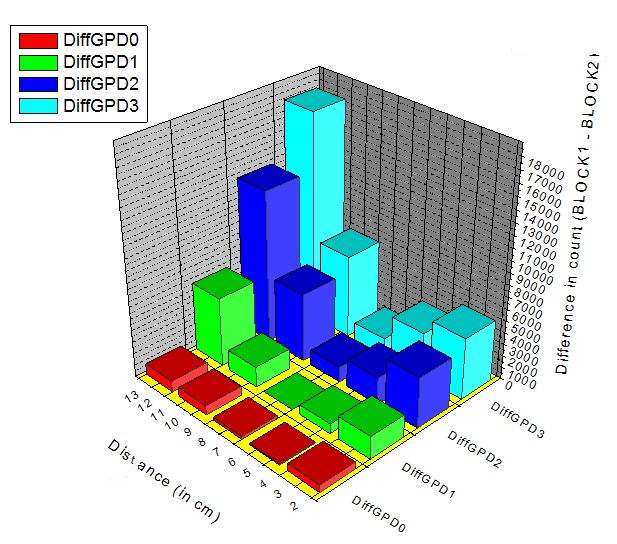}
\mbox{\textbf{Fig.4.} The 2D bar graph of difference Compton photo peak counts }
\end{center}
\end{figure}

\begin{figure}[H]
\begin{center}
\includegraphics[scale=0.7]{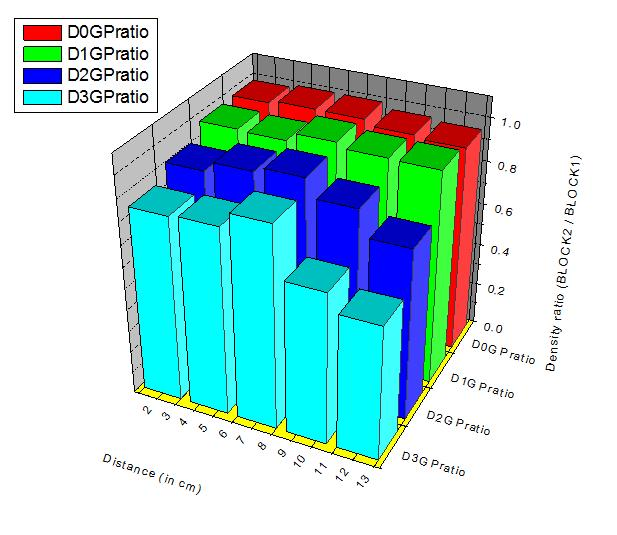}
\mbox{\textbf{Fig.5.} The density ratio as a function of lateral distance}
\end{center}
\end{figure}

\section{\textbf{References}}
\begin{enumerate}
	\item Lawson, L., \"Backscatter Imaging\", Materials Evaluation, 60 (11), 1295-1316, 2002
	\item Harding G, Inelastic photon scattering: Effects and applications in biomedical science and industry Radiation Phys. Chem.1997, 50,91-111
	\item Hussein E M A Whynot T M, A compton scattering method for inspecting concrete 
structures Nucl. Instr. Method 1989,A283,100-106
  \item P Zhu, P Duvauchelle, G Peix and D Babot: X-ray Compton backscattering techniques for process tomography: imaging and charecterization of materials; Meas. Sci. Technol. 7.
  \item C.F.Poranski, E.C.Greenawald and Y.S.Ham; X-Ray Backscatter Tomography: NDT Potential and Limitations; Materials Science Forum Vols. 210-213 (1996) pp. 211-218.
  \item R.Cesareo, F.Balogun, A.Brunetti, C.Cappio Borlino, $90^o$ Compton and Rayleigh measurements and imaging; Radiation Physics and Chemistry 61 (2001) 339-342. 
\end{enumerate}

\section{Acknowledgements}
\begin{itemize}
  \item{Dr. N.Mohankumar, Head, Radiological Safety Division,IGCAR, Kalpakkam.}
  \item{Ramar, IGCAR.}
  \item{Priyada, IGCAR.}
\end{itemize}  
\end{document}